\documentclass{iopart}
\usepackage{graphicx}
\usepackage{dcolumn}
\usepackage{bm}
\usepackage{latexsym}

\usepackage{color}

\usepackage{times}
\definecolor{darkblue}{rgb}{0, 0, 0.8}
\usepackage[colorlinks=true, breaklinks=true, linkcolor=darkblue, citecolor=darkblue, urlcolor=darkblue]{hyperref}
\newcommand{\doilink}[2]{\href{http://dx.doi.org/#1}{#2}}

\begin{document}
\title{Measuring the eccentricity of the Earth orbit with a nail and a piece of plywood}
\author{Thierry Lahaye$^{1,2}$\footnote{Present address: Laboratoire Charles Fabry, CNRS UMR 8501, Institut d'Optique, F-91127 Palaiseau cedex, France.}}
\address{$^1$ Universit\'e de Toulouse, UPS, Laboratoire Collisions Agr\'egats R\'eactivit\'e, IRSAMC ; F-31062 Toulouse, France}
\address{$^2$ CNRS, UMR 5589, F-31062 Toulouse, France}
\date{\today}

\begin{abstract}
I describe how to obtain a rather good experimental determination of the eccentricity of the Earth orbit, as well as the obliquity of the Earth rotation axis, by measuring, over the course of a year, the elevation of the Sun as a function of time during a day. With a very simple ``instrument'' consisting of an elementary sundial, first-year students can carry out an appealing measurement programme, learn important concepts in experimental physics, see concrete applications of kinematics and changes of reference frames, and benefit from a hands-on introduction to astronomy.
\end{abstract}

\maketitle

\section{Introduction}


One of the cornerstones of introductory courses in classical mechanics is the derivation of Kepler's laws. In particular, the derivation of Kepler's first law, stating that the trajectory of a planet is an ellipse with the Sun located at one of the foci, is an important application of Newton's laws to a multidimensional problem. However, very few students are aware of the fact that the eccentricities of the planets of the Solar system are actually quite small, with trajectories very close to a circle, which makes Kepler's achievement (based on Tycho Brahe's measurements) even more remarkable.

Here, I describe a simple measurement programme, suitable for first-year university students, consisting in measuring the elevation of the Sun as a function of time during a day, and in repeating this typically once a week over a full year. By measuring the maximal elevation $h_{\max}$ of the Sun, and the time $t_{\max}$ at which this maximum occurs (i.e. the \emph{true local noon}), students can readily check that these quantities vary a lot over the year. The change in $h_{\max}$ is essentially related to the obliquity $\varepsilon$ of the Earth over the ecliptic, and thus allows for quite an accurate determination of $\varepsilon$ (as well as that of the latitude of observation). The change of $t_{\max}$ over a year gives an experimental determination of the \emph{equation of time} $E(t)$, i.e. the difference between the mean local noon and the true local noon, and allows for a determination of the eccentricity $e$ of the Earth orbit~\cite{rees1991}. This is a rewarding result for students to realize that with such simple measurements they can obtain good experimental values for the above quantities, and that with careful observations one can perform `science without instruments' as did the astronomers of various antique civilizations~\cite{gutzwiller1998,krisciunas2010}.

This article is organized as follows. I first describe how to measure in a simple way the elevation of the Sun versus time over a day, with an accuracy of about $1^\circ$. Then I give the results I obtained for $h_{\max}(t)$ and $E(t)$ by repeating the measurement about once a week for one year, starting in August 2010. I show how one can extract the obliquity $\varepsilon$ of the Earth's axis and the eccentricity $e$ of its orbit by fitting the experimental data with simple, analytic expressions. Finally, possible extensions of the work are proposed. \ref{append:a} contains a brief reminder on basic notions of spherical astronomy, and should be read first by readers not familiar with these notions. In the remaining appendices, the derivation of the analytic expressions used for fitting the data is given, so that the article is self-contained.

\section{Measurements}

We are interested in studying the motion of the Earth around the Sun. Using the relativity of motion, we can thus simply measure the apparent motion of the Sun on the celestial sphere, i.e. the time-dependance of two angles that define the position of the Sun in the sky.

\begin{figure}[t]
\centerline{\includegraphics[width=70mm]{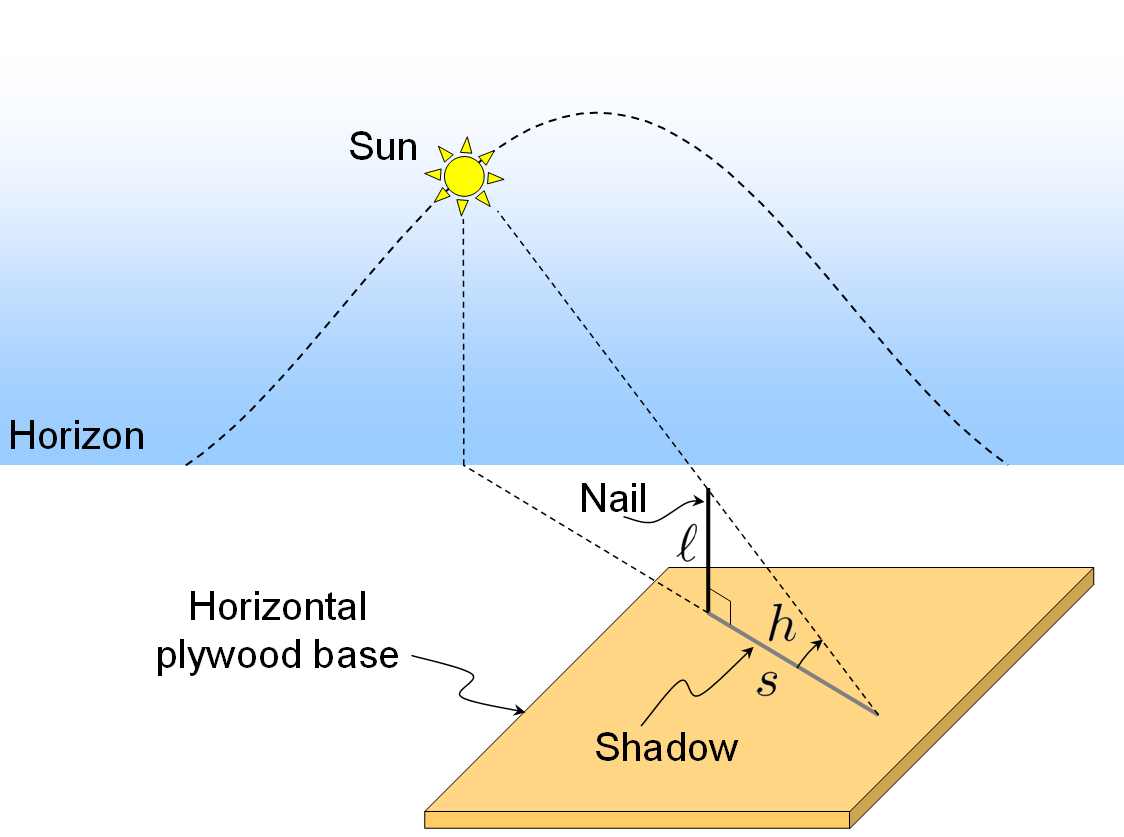}}
\caption{Schematic view of the elementary sundial.}
\label{fig:Setup}
\end{figure}

As we shall see, for our purpose, it is sufficient to measure the elevation of the Sun (also called \emph{altitude}, or \emph{height}) above the horizon, i.e. the angle $h$ shown in Fig.~\ref{fig:Setup}. This can be done very simply by measuring the length $s$ of the shadow of a vertical \emph{gnomon} (i.e. a rod with a sharp point) of length $\ell$. Then the elevation of the Sun is given by $h=\arctan(\ell/s)$.

Contrary to the case where one would measure also the azimuthal position of the Sun, here, the orientation of the horizontal base does not need to be fixed. One of the advantages of using such a simple setup is therefore that one can change the position of the sundial over the course of the day, e.g. in order to operate indoor.

\subsection{Construction and use of an elementary sundial}

In practice, I used as a gnomon a steel nail protruding from a plywood base of size $20\times20$~cm$^2$. In order to have the nail as orthogonal to the base as possible, a hole with a diameter slightly less than that of the nail was first drilled into the plate using a drill press. The nail I used has a length $\ell=69$~mm above the plate. To measure $h$, one simply installs this elementary sundial on a horizontal surface in the sunlight, and measures with a ruler the length of the shadow. Two effects limit the accuracy of the measurement: first, due to the final angular diameter of the Sun, the shadow is slightly blurred; second, the horizontality of the base when installed on the floor of a room, or on a table, is not perfect~\cite{horizontality}. In practice, an accuracy of typically one degree is easily obtained (this can be estimated quantitatively by repeating the measurement several times with the sundial in different positions, in a short interval over which $h$ barely varies, and observing the dispersion of the results).

Concerning the determination of the time $t$ at which $h(t)$ is measured, an accuracy of one minute is sufficient for our purpose, and thus a simple wristwatch can be used. However it is wise to check that the watch indicates the correct time before starting a series of measurements. Nowadays, this can be done very easily using the websites of national time agencies~\cite{webnist} that give access to the legal time with an accuracy of one second or better.

\subsection{Measuring the altitude of the Sun over a day}

\begin{figure}[t]
\centerline{\includegraphics[width=75mm]{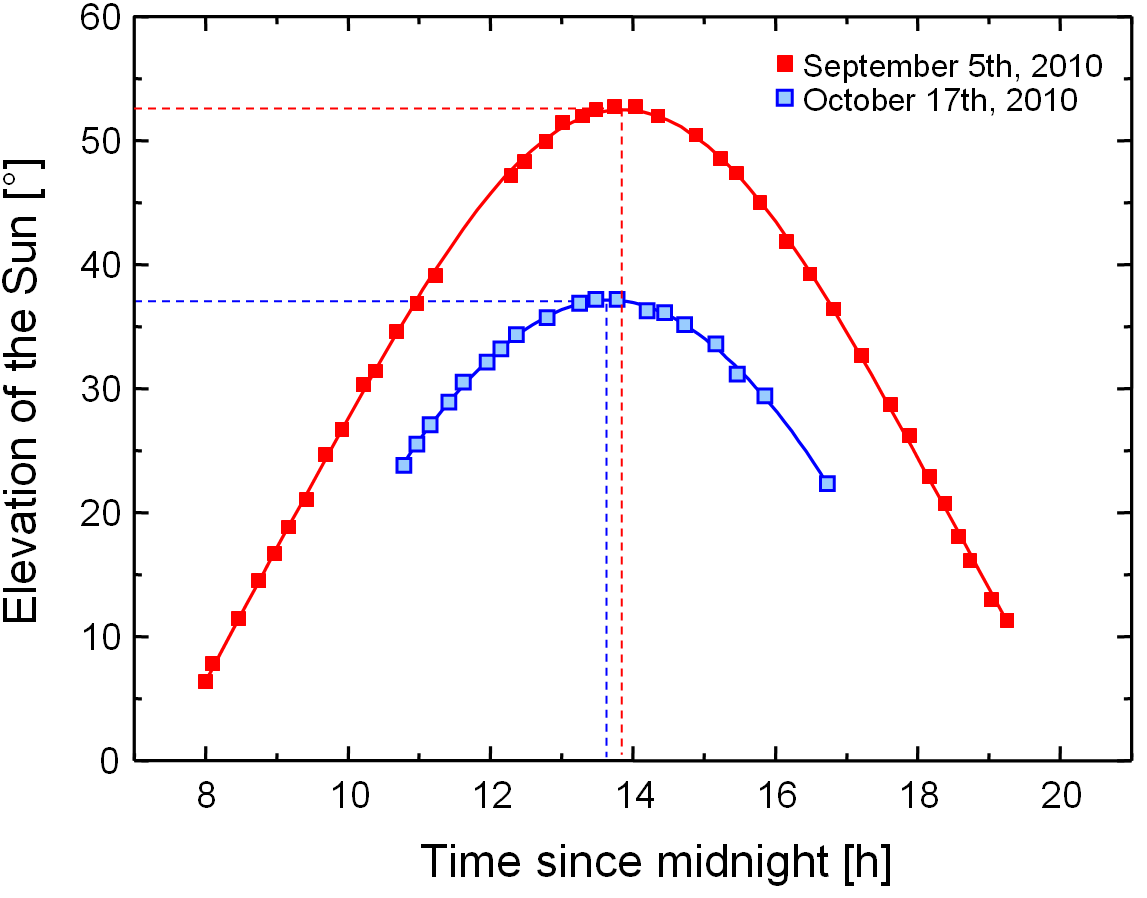}}
\caption{Measured elevation of the Sun over the course of a day, for two different dates. The solid lines are fits by a polynomial (see text). The dashed lines display the values of $h_{\rm max}$ and $t_{\rm max}$.}
\label{fig:Elev}
\end{figure}

Figure~\ref{fig:Elev} shows two measurements of $h(t)$, where $t$ is the legal time, performed in Toulouse, France (latitude $\varphi=43^\circ\,36'$~N, longitude $\lambda=1^\circ\,27'$~E) at two different dates. It is very clear from the data that the maximal height $h_{\max}$ of the Sun depends on the date; this is in general well known as it is related to the cycle of seasons.

However, what appears also clearly on Fig.~\ref{fig:Elev} is that the time $t_{\max}$ at which this maximum occurs also depends on the date; this however is not widely known by the students, nor even by some physicists, probably because the effect is relatively small (a few minutes) though perfectly measurable even with our crude setup.

In order to proceed, we need to extract from $h(t)$ the two quantities $h_{\max}$ and $t_{\max}$. The theoretical expression of $h(t)$ is derived in \ref{sec:TheorElev}; however we can at this stage keep an empiric approach and just fit the data with a simple function. As $h(t)$ is symmetric about $t=t_{\max}$ (provided one neglects the motion of the Sun with respect to the fixed stars over a few hours, which is reasonable given the accuracy of our measurements), I chose to fit the data with the following polynomial:
\begin{equation}
h(t)=h_{\max}+\sum_{i=1}^{3}h_{2i}(t-t_{\max})^{2i},
\end{equation}
where the five adjustable parameters are $h_{\max}$, $t_{\max}$, and the coefficients $h_{2,4,6}$. I chose to go up to sixth order in order to get a nicer fit at small elevations (in the mornings and evenings) but if the data is taken only for a few hours around $t_{\max}$ ($\pm 3$ to 4~hours around $t_{\max}$ are enough to determine the quantities of interest) one can use only a fourth-order polynomial without affecting the results. Such fits are shown as solid lines in Fig.~\ref{fig:Elev}.

The accuracy in the determination of $h_{\rm max}$ and $t_{\rm max}$ obviously depends on the number of measurement points; for the data presented in Fig.~\ref{fig:Elev}, they are respectively of about $0.2^\circ$ and $1$~min, as data points were collected for several hours before and after $t_{\rm max}$, at a rate of four points per hour typically. When the weather is partly cloudy, one sometimes has to stop taking data for a while, and the accuracy in the determination of $h_{\rm max}$ and $t_{\rm max}$ is thus not as good.

\subsection{Annual variation of $h_{\rm max}$: determination of $\varphi$ and $\varepsilon$ }

\begin{figure}[t]
\centerline{\includegraphics[width=75mm]{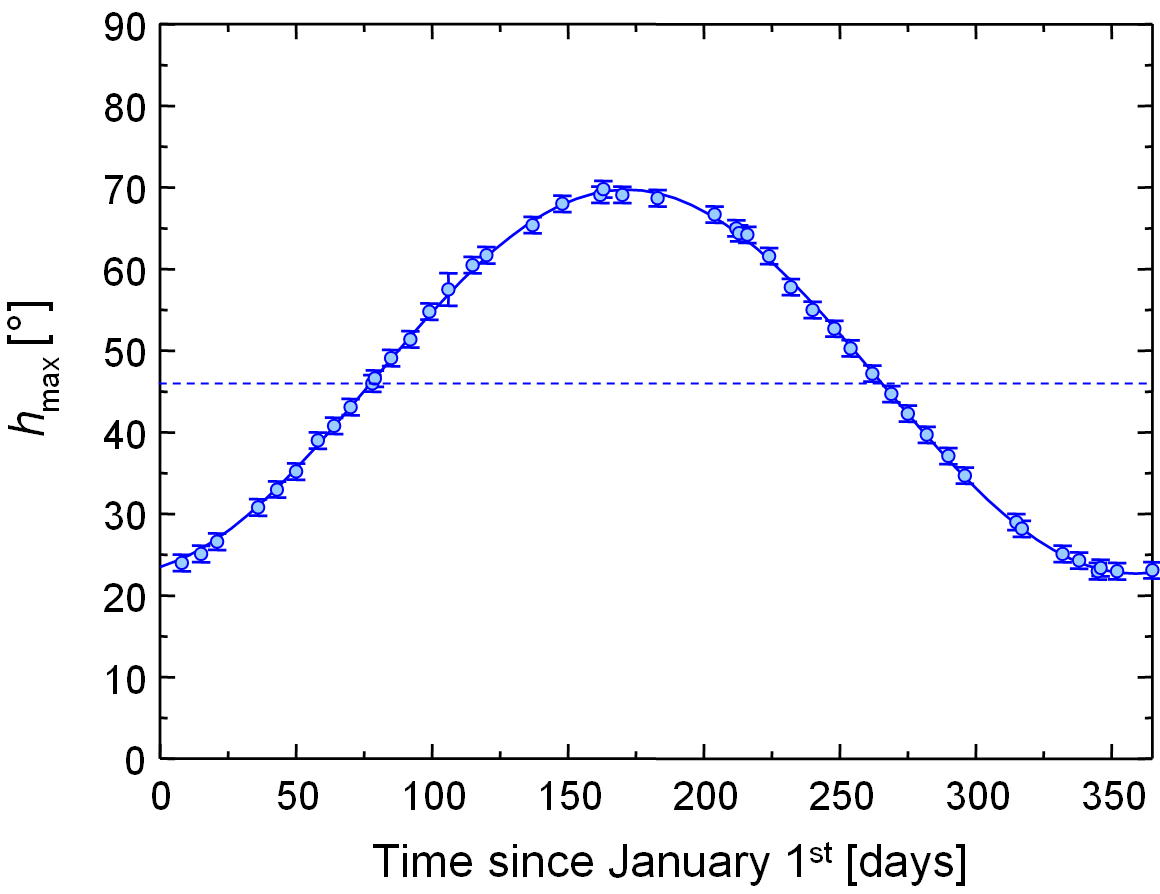}}
\caption{Measured maximal elevation of the Sun (points with error bars). We observe a sinusoidal variation around $\pi/2-\varphi$, with and amplitude of $2\varepsilon$. The solid line is a fit to the simple model discussed in the text.}
\label{fig:MaxElev}
\end{figure}

By repeating the above measurements typically once per week for a year, the annual variations of $h_{\rm max}$ and $t_{\max}$ can be obtained. Figure~\ref{fig:MaxElev} shows the maximal elevation $h_{\rm max}$ of the Sun as a function of time. One observes a quasi-sinusoidal variation with a period of one year.

The qualitative explanation for this phenomenon is simple, and is usually part of geography courses in elementary education, but I repeat it here for completeness. Due to the obliquity $\varepsilon$ of the Earth rotation axis, the angle between the Earth axis and the line joining the center of the Earth to the Sun varies between $\pi/2-\varepsilon$ (at the summer solstice, when the Earth axis leans towards the Sun) and $\pi/2+\varepsilon$ (at the winter solstice). Correspondingly, the angular distance between the Sun and the celestial pole varies with a period of one year, and an amplitude of $\varepsilon$. This induces a similar variation of $h_{\rm max}$, around a mean value which depends on the observer's latitude. For a quantitative treatment, the reader is referred to~\ref{sec:TheorMaxElev}.

\subsection{Annual variation of $t_{\rm max}$: determination of $e$}

We now turn to a more subtle measurement concerning the variation of $t_{\rm max}$, which defines the true local noon. It is convenient to convert the measured values into a quantity called the \emph{equation of time}, that we shall denote by $E$, defined as the difference between the mean local noon and the true local noon (our measured $t_{\max}$). The former is obtained from the legal noon, given by clocks (corrected if necessary by one hour in summer due to daylight saving time) by adding (subtracting) four minutes for each degree of longitude west (east) from the reference meridian of the corresponding time zone. For instance, in Toulouse (longitude $\lambda=1^\circ\,27'$~E), one needs to subtract $t_{\max}$ from 13.00~h in winter time and 14.00~h in summer time, and then subtract another 5.8~min to correct for the longitude, to obtain the equation of time $E$. For instance, on October 17th (see Fig.~\ref{fig:Elev}), we have $t_{\max}=13.66$~h, thus giving $E=60(14-13.66)-5.8=14.6$~min.

\begin{figure}[t]
\centerline{\includegraphics[width=70mm]{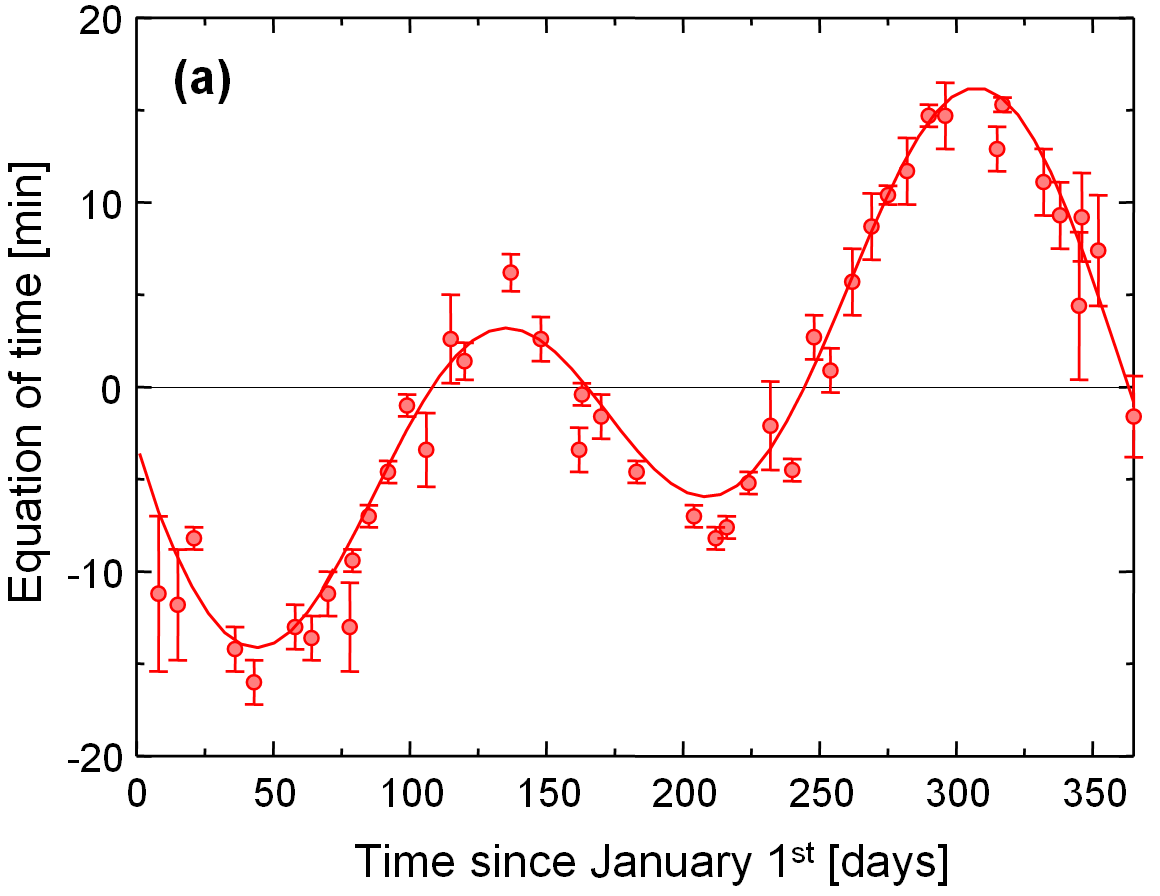}\hfill
\includegraphics[width=70mm]{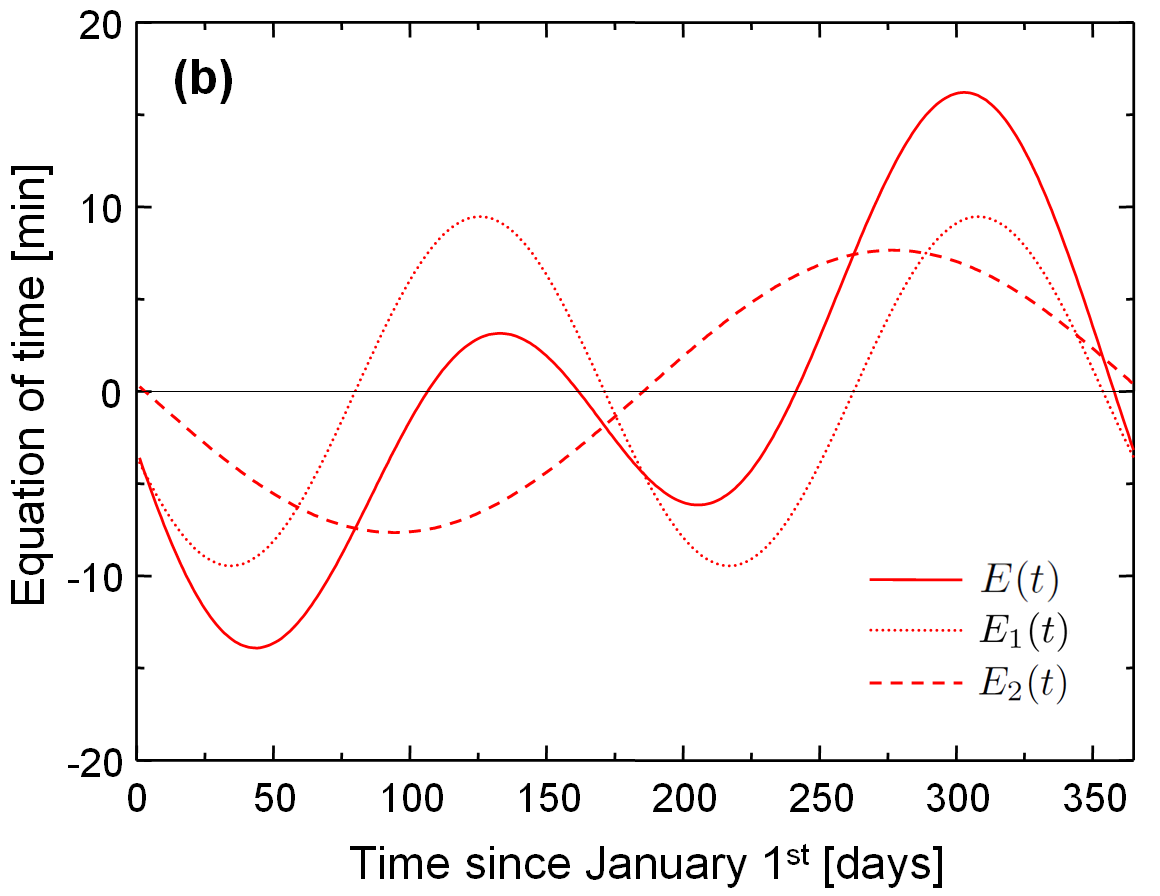}}
\caption{(a) Measured equation of time $E$ (points with error bars). The solid line is a fit to the simple model of the equation of time given in the appendix. (b) Combining the effect of the Earth obliquity ($E_1(t)$, dotted line) and of the eccentricity of the Earth orbit ($E_2(t)$, dashed line) gives the full expression of the equation of time $E(t)$ (solid line).}
\label{fig:EOT}
\end{figure}

Figure~\ref{fig:EOT} gives the results obtained by measuring $E(t)$ over a year. One observes a non-trivial behavior, the equation of time varying between a maximum of about 16~min in autumn and a minimum of about $-15$~min in winter, and vanishing at four different dates.

Physically, the origin of the equation of time lies in the fact that the duration of the true solar day, i.e. the time elapsed between two successive transits of the Sun across the observer's meridian, is not constant over a year. The solar day would have a constant duration if, along the year, the Sun moved on the celestial sphere (i) at constant angular velocity, and (ii) along the celestial equator (this defines the so-called \emph{mean Sun}; the time between two transits of the mean Sun defines the mean solar day of 86,400~s). However, these two assumptions are both wrong: since the Sun moves along the ecliptic, which is inclined with respect to the equator due to the obliquity $\varepsilon$ of the Earth axis, the motion of its projection on the equator is irregular (it coincides with the mean Sun at the time of the equinox, then lags behind the mean Sun for a quarter of the year, catches up at the solstice, and then is ahead of the mean Sun for another three months). This contribution $E_1$ to the equation of time thus has a 6-month period. Moreover, via Kepler's second law of areal velocity (see \ref{sec:TheorEOT}) the angular velocity of the apparent motion of the Sun is not constant over the year, due to the fact that the Earth orbit is not circular: for instance, when the Earth--Sun distance is smaller (in January), the Sun moves faster along the  ecliptic. This contribution $E_2$ to $E(t)$ has obviously a one-year period. Combining these two contributions explains the temporal variation of the equation of time (see Fig.~\ref{fig:EOT}b; \ref{sec:TheorEOT} gives the derivation of the analytical expressions of $E_1$ and $E_2$).

\section{Exploiting the data}

\subsection{Obliquity of the Earth}

It is easy to show (see \ref{sec:TheorMaxElev}) that to a very good approximation, one has:
\begin{equation}
h_{\rm max}=\frac{\pi}{2}-\varphi+\varepsilon\sin\left(\frac{2\pi}{T}(t-t_0)\right),
\label{eq:hmax}
\end{equation}
where $\varphi$ is the latitude of the place of observation, $\varepsilon$ the obliquity of the Earth axis, $T$ the duration of the year, and $t_0$ the date of the vernal equinox.
When fitting the data by equation~(\ref{eq:hmax}) with the four previous quantities as adjustable parameters, we obtain
\begin{equation}
\left\{
\begin{array}{l}
\varphi=43.8\pm 0.2^\circ,\\
\varepsilon =23.5\pm 0.1^\circ,\\
T=374\pm 6\;{\rm d},\\
t_0=78\pm 1 \;{\rm d},{\rm \;i.e.\;March\;19^{th}},
\end{array}
\right.
\end{equation}
which is close to the accepted values (respectively, $43.60^\circ$, $23.44^\circ$, $365.25$~d, and March 20$^{\rm th}$.) Note that by repeating the measurements over the course of several years, a much more accurate determination of the duration $T$ of the year could be achieved.

\subsection{Eccentricity of the Earth orbit}

We show in \ref{sec:TheorEOT} that a good approximation of the equation of time is given by:
\begin{equation}
E(t)=\frac{d}{2\pi}\left[\frac{1-\cos{\varepsilon}}{2}\sin\left(\frac{4\pi}{T}(t-t_0)\right)-2e\sin\left(\frac{2\pi}{T}(t-t_1)\right)\right]
\label{eq:EOT}
\end{equation}
where $d$ is the duration of a day (i.e. 1440 minutes), $e$ the eccentricity of the Earth orbit, and $t_1$ the date of perihelion passage.

Fitting the data shown in Fig.~\ref{fig:EOT} by equation~(\ref{eq:EOT}) with $e$ and $t_1$ as adjustable parameters (and using the values determined above for $\varepsilon$ and $T$), we obtain
\begin{equation}
\left\{
\begin{array}{l}
e=0.017\pm 0.001,\\
t_1=1\pm 5\;{\rm d}, {\rm\; i.e.\;January\;1^{st}},
\end{array}
\right.
\end{equation}
again in relatively good agreement with the values $e=0.0167$ and $t_1=3$~d found in the literature.

\section{Conclusion and outlook}
I have shown that with very modest equipment, one can measure with reasonable accuracy some of the orbital elements of the Earth, and in particular its eccentricity, despite its relatively small value. The above measurements can be the basis of further activities for students. Among them, one can list the following ones, given here under the form of exercises:
\begin{itemize}
\item Use Eq.~(\ref{eq:elev}) of appendix~\ref{sec:TheorElev} to calculate the length of daytime as a function of the latitude along the year and compare it to the one obtained from the ephemerides given in calendars.
\item Show that the duration of a solar day is $(1+{\rm d}E/{\rm d}t)\times 86,\!400$~s. What are its minimal and maximal values?
\item Using a sundial with a fixed base, check experimentally that the azimuthal position of the Sun when reaches its highest elevation is always the same (i.e., South) throughout the year~\cite{footnote}.
\item Still with a fixed-base sundial, plot experimentally the curve traced out over the year by the end of the shadow at the \emph{mean} local noon. This eight-shaped curve is called an \emph{analema} and is sometimes encountered on sundials in order to allow for a computation of the legal time from the measured solar time.
\end{itemize}

\section*{Acknowledgements}
I thank Maxime Lahaye and Irina Lahaye for help in taking the data, Giovanni Luca Gattobigio for discussions, and Lucas B\'eguin, Antoine Browaeys and Renaud Mathevet for useful suggestions that helped me improve the manuscript.

\appendix

\section{A quick reminder on spherical astronomy}
\label{append:a}

\begin{figure}[t]
\centerline{\includegraphics[width=100mm]{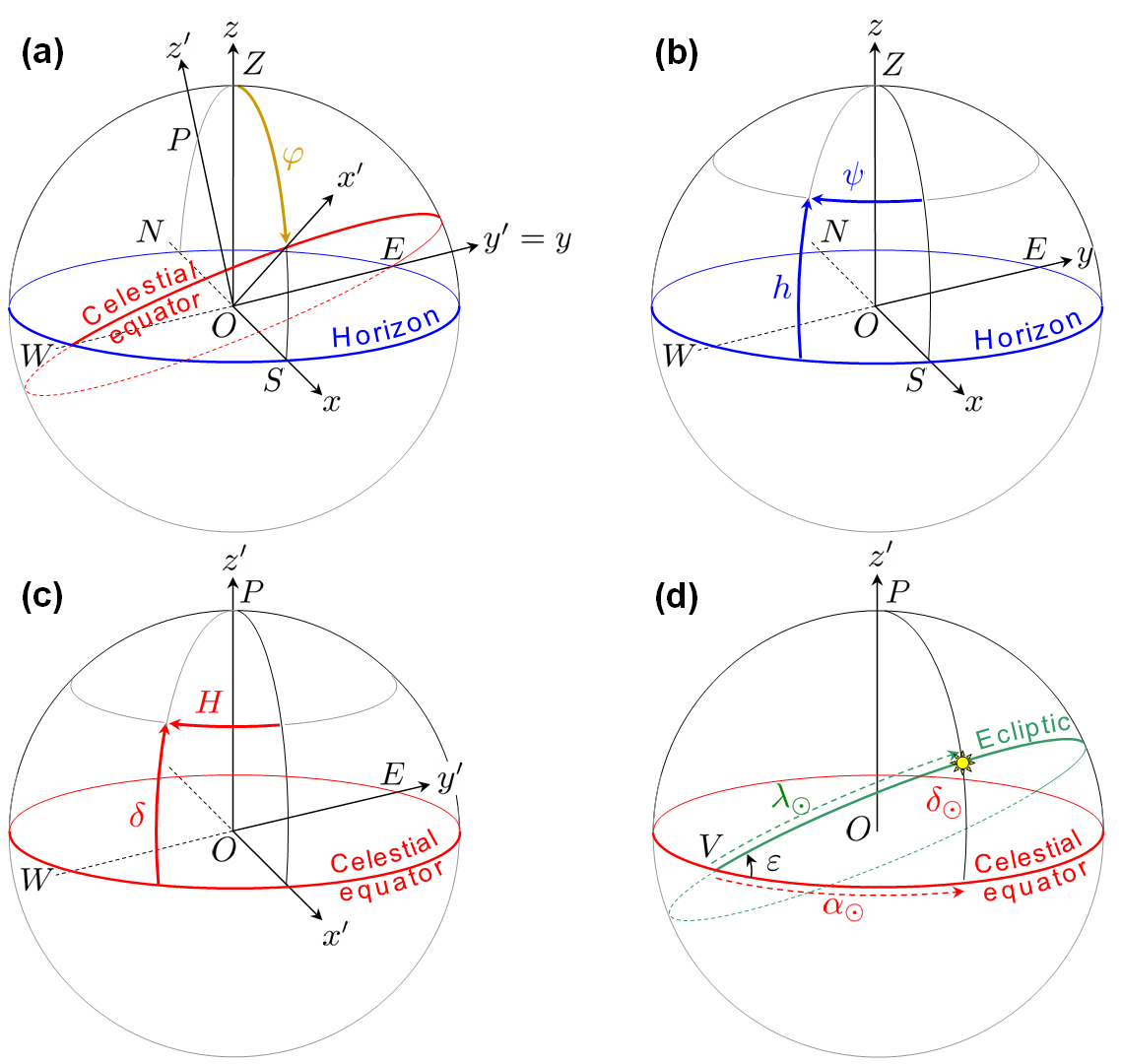}}
\caption{Geometry of the celestial sphere. (a): important points, axes and circles on the celestial sphere. (b): definition of the coordinates in the horizon system. (c): definition of the coordinates in the equatorial system. (d): definition of the coordinates of the Sun $\odot$ in the equatorial system during its motion along the ecliptic.}
\label{fig:celestial:sphere}
\end{figure}

This section consists in a minimalist reminder about basic terms and notions of spherical astronomy needed for the understanding of the paper. The reader is referred to the first sections of Ref.~\cite{gutzwiller1998} for a similar but more exhaustive reminder. A very detailed and accessible introduction to spherical astronomy can be found online in the celestial mechanics lecture notes of Ref.~\cite{tatum}, or in standard textbooks about spherical astronomy~\cite{textbook}.

To an Earth-bound observer $O$, celestial bodies appear to move on a sphere centered on himself, the \emph{celestial sphere} (Fig.~\ref{fig:celestial:sphere}a). The local vertical points toward the \emph{zenith} $Z$; the great circle perpendicular to the vertical is the \emph{horizon}. Over a day, ``fixed'' stars appear to rotate around the \emph{north celestial pole} $P$ (close to the star \emph{Polaris}). The great circle going through $Z$ and $P$ crosses the horizon in two points defining the South $S$ and the North $N$; the two other cardinal points on the horizon (East $E$ and West $W$) are deduced from $S$ and $N$ by a $90^\circ$ rotation around the vertical axis.

The position of celestial bodies can be specified in spherical coordinates by two angles, once a reference system has been chosen. Two reference systems are particularly useful: the \emph{horizon system}, in which measurements are made in practice, and which is dependent on the observer's location on Earth, and the \emph{equatorial system}, which is defined by the directions of ``fixed'' stars.
\begin{itemize}
\item
In the horizon system (Fig.~\ref{fig:celestial:sphere}b), the local vertical, pointing towards the \emph{zenith}, is chosen as the $z$-axis. The position of a point on the celestial sphere is specified using the \emph{altitude} (or \emph{elevation}) $h$ (angle between the radius-vector of the point and the horizon), and the \emph{azimuth $\psi$}, counted along the horizon, starting from the South and counted positive towards the West (note that other conventions exist for the choice of the azimuth origin).
\item
In the equatorial system (Fig.~\ref{fig:celestial:sphere}c), $OP$ is chosen as the polar axis. The intersection of the plane perpendicular to $OP$ with the celestial sphere defines the \emph{celestial equator}. The position of a point on the celestial sphere is given by the declination $\delta$ (angle from the celestial equator to the body) and the \emph{hour angle} $H$ (counted along the celestial equator, from the south to the equatorial projection of the body).
\end{itemize}
The equatorial system is obtained by rotating the horizon system around $OE$ by an angle $\pi/2-\varphi$, where $\varphi$ is the (geographical) latitude of the point of observation; for instance, in at the North pole ($\varphi=\pi/2$) the equatorial and horizon systems coincide.

Finally, the \emph{ecliptic} is the great circle along which the apparent annual motion of the Sun (traditionally denoted by the astronomical symbol $\odot$) takes place on the celestial sphere. It is inclined on the celestial equator by the obliquity $\varepsilon$ of the Earth axis.

The angular position of the Sun along the ecliptic is given by the ecliptic longitude $\lambda_\odot$, whose origin is taken at the vernal point $V$ (the point where the ecliptic crosses the celestial equator, and where the Sun is located at the time of the spring equinox in the Northern hemisphere). The angular distance on the celestial equator between $V$ and the projection of the Sun on the equator is the Sun's \emph{right ascension} $\alpha_\odot$.

\section{Expression of $h(t)$ over a day}\label{sec:TheorElev}

From the definitions given above, in the equatorial coordinate system $(x',y',z')$, the coordinates of the Sun read:
\begin{equation}
S'=\left(
\begin{array}{l}
x'=\cos\delta\,\cos H\\
y'=-\cos\delta\,\sin H\\
z'=\sin\delta
\end{array}
\right),
\end{equation}
with $\delta$ the declination and $H$ the hour angle of the Sun. In the horizon system $(x,y,z)$, they read
\begin{equation}
S=\left(
\begin{array}{l}
x= \cos h\,\cos \psi\\
y=-\cos h \,\sin \psi\\
z=\sin h
\end{array}
\right),
\end{equation}
where $h$ is the altitude of the Sun, and $\psi$ its azimuth. Since the equatorial system is deduced from the horizon system by a rotation of angle $\pi/2-\varphi$ around the $(Oy)=(Oy')$ axis, the $(x,y,z)$ coordinates are obtained by multiplying the $(x',y',z')$ ones by the following rotation matrix:
\begin{equation}
R=\left(
\begin{array}{ccc}
\sin\varphi & 0 & -\cos\varphi \\
0 & 1 & 0 \\
\cos\varphi & 0 & \sin \varphi \\
\end{array}
\right).
\end{equation}
From the last component of the relation $S=RS'$ we get
\begin{equation}
\sin h = \cos\varphi\,\cos\delta\,\cos H+\sin\varphi\,\sin\delta
\label{eq:elev}
\end{equation}
which gives the elevation of the Sun as a function of time (i.e. the hour angle $H$) for given location and declination of the Sun.

\section{A simple model for $h_{\max}(t)$}\label{sec:TheorMaxElev}

From equation (\ref{eq:elev}) above, we find immediately that the elevation of the Sun becomes maximal when  $\cos H=1$ and reaches the value $h_{\rm max}$ fulfilling
\begin{equation}
\sin h_{\rm max}= \cos\varphi\,\cos\delta+\sin\varphi\,\sin\delta=\cos(\varphi-\delta)
\end{equation}
whence
\begin{equation}
h_{\rm max}=\frac{\pi}{2}-\varphi+\delta.
\label{eq:hmax:app}
\end{equation}
We now need to express the time dependance of the declination of the Sun. Using Fig.~\ref{fig:celestial:sphere}, one can show (see Eq.~(\ref{eq:sindelta}) below) that $\sin\delta_\odot=\sin\varepsilon\sin\lambda_\odot$, which can be simplified to $\delta_\odot\simeq\varepsilon\sin\lambda_\odot$ to a very good approximation (even though $\varepsilon\simeq23^\circ$, the maximal difference is smaller than $0.3^\circ$, i.e. negligible in comparison with our experimental uncertainties). Making the further simplification that the solar ecliptic longitude $\lambda_\odot$ increases linearly in time (i.e. assuming here that the eccentricity of the Earth orbit is $e=0$), we have $\lambda_\odot=2\pi(t-t_0)/T$, with $t_0$ the date of spring equinox. Combining this simple sinusoidal approximation for $\delta_\odot(t)$ and Eq.~(\ref{eq:hmax:app}) finally yields Eq.~(\ref{eq:hmax}) of the main text.

\section{A simple model for the equation of time}\label{sec:TheorEOT}

Following e.g. Ref.~\cite{rees1991}, a good approximation of the theoretical expression of the equation of time can be obtained in the following way. If the eccentricity of the Earth orbit were $e=0$, and if the obliquity of the Earth were $\varepsilon=0$, one would have $E=0$. We can thus expect that by calculating separately the small contributions $E_1$ and $E_2$ of both the obliquity and of the eccentricity, and adding them, a good approximation of $E$ is obtained: one basically expands $E$ to the lowest orders in $e$ and $\varepsilon$.

We first calculate the contribution $E_1$ to $E$ arising from the nonzero value of the obliquity. Here, we can neglect the ellipticity of the Earth orbit, and assume that the motion of the Sun around the Earth takes place on a circle, and thus, using Kepler's second law, at a constant angular velocity. The longitude $\lambda_\odot$ of the Sun along the ecliptic (see Fig.~\ref{fig:celestial:sphere}d) thus increases linearly in time as
\begin{equation}
\lambda_\odot=\frac{2\pi}{T}(t-t_0),
\label{eq:lambdat}
\end{equation}
where $T$ is the length of a year and $t_0$ the date of the vernal (i.e. spring) equinox. However the mean Sun is a fictitious body that moves at constant angular velocity along the \emph{celestial equator}, not the ecliptic. The difference between $\lambda_\odot$ and $\alpha_\odot$, once converted to time via the correspondance $1^\circ\leftrightarrow 4\;{\rm min}$, thus gives the obliquity contribution to the equation of time:
\begin{equation}
E_1=(4\;{\rm min}/^\circ)\times(\lambda_\odot-\alpha_\odot)
\end{equation}
(for readability, from now on I shall drop the subscript $\odot$). We thus have to express $\alpha$ as a function of $\lambda$, and then use (\ref{eq:lambdat}), to obtain $E_1(t)$. To find the relation between $\alpha$ and $\lambda$, we introduce the \emph{equatorial} frame $(Oxyz)$ with origin at the center of the celestial sphere, the $z$ axis pointing towards the celestial pole, and the $x$ axis towards the vernal point $V$, and another, \emph{ecliptic} frame $(Ox'y'z')$ obtained from the former by a rotation of angle $\varepsilon$ around $Ox$. The coordinates of the Sun in the equatorial frame are $(\cos\delta\cos\alpha,\cos\delta\sin\alpha,\sin\delta)$, and in the ecliptic frame $(\cos\lambda,\sin\lambda,0)$. Since the rotation matrix from the equatorial to the ecliptic frame reads
\begin{equation}
\left(
\begin{array}{ccc}
1 & 0 & 0 \\
0 & \cos\varepsilon & \sin\varepsilon \\
0 & -\sin\varepsilon & \cos\varepsilon
\end{array}
\right),
\end{equation}
one can relate $(\alpha,\delta)$ to $\lambda$. We obtain:
\begin{eqnarray}
\tan\alpha&=&\tan\lambda\cos\varepsilon, \label{eq:tan}\\
\sin\delta&=&\sin\varepsilon\sin\lambda. \label{eq:sindelta}
\end{eqnarray}
Therefore, using (\ref{eq:tan}):
\begin{equation}
\lambda-\alpha=\lambda-\arctan\left(\cos\varepsilon\tan\lambda\right).
\end{equation}
From this expression it may not obvious to see the time dependence (in particular the periodicity) of $E_1$. We can obtain a better understanding (and a convenient expression) by noticing that $\cos\varepsilon$ is actually close to one (for $\varepsilon=23.44^\circ$, we have $\cos\varepsilon\simeq0.9174$). If one Taylor-expands $f(x,A)=x-\arctan(A\tan x)$ around $A=1$, one gets $f(x,A)\simeq (1-A)\sin(2x)/2$, and thus, putting everything together, we obtain our final expression for $E_1$:
\begin{equation}
E_1(t)\simeq\frac{d}{2\pi}\frac{1-\cos{\varepsilon}}{2}\sin\left(\frac{4\pi}{T}(t-t_0)\right).
\label{eq:e1}
\end{equation}
where $d$ is the duration of a day. The dotted line on Fig.~\ref{fig:EOT}b shows $E_1(t)$.

\begin{figure}[t]
\centerline{\includegraphics[width=45mm]{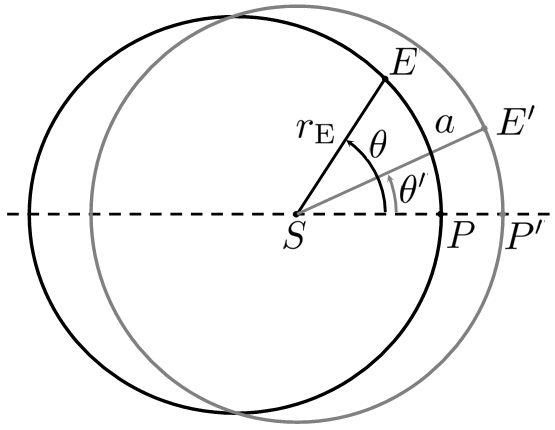}}
\caption{Circular (gray) and elliptical (black) orbits around the Sun $S$ with the same semi major axes $a$. The ellipticity of the ellipse is $e=0.3$ for clarity.}
\label{fig:ellipses}
\end{figure}

We now turn to the calculation of $E_2$, the contribution of the eccentricity of the Earth orbit: since its orbit is elliptic, the Earth does not move at constant angular velocity along its orbit, and the difference in angular position between the Earth and a fictitious body moving at constant speed with the same period gives the contribution $E_2$ to the equation of time. Figure~\ref{fig:ellipses} shows the trajectory of the Earth $E$ (black ellipse with the Sun $S$ at one focus) and of the fictitious Earth $E'$ having a circular orbit centered on $S$, with the same orbital period as $E$. From Kepler's third law $T^2/a^3=4\pi^2/(GM_{\rm Sun})$, the radius $r$ of the circular orbit is thus equal to the semi-major axis $a$ of the elliptical orbit of $E$. Using the perihelia $P$, $P'$ as the origins of angles, the polar angles defining the positions of $P$ and $P'$ are $\theta$ and $\theta'$, respectively. We are interested in finding the difference in angular positions $\vartheta\equiv\theta'-\theta$, as a function of $\theta$, and then as a function of time. We will perform the calculation by keeping only first-order terms in the eccentricity $e$.

The equation of the elliptical orbit of $E$ reads~\cite{goldstein}
\begin{equation}
r_{\rm E}=\frac{a(1-e^2)}{1+e\cos\theta}.
\end{equation}
Moreover, Kepler's second law about the areal velocity implies that
\begin{equation}
\frac{1}{2}a^2\dot{\theta}'=\frac{\pi a^2}{T}
\end{equation}
for the circular orbit, and
\begin{equation}
\frac{1}{2}r_{\rm E}^2\dot{\theta}'=\frac{S_{\rm ellipse}}{T}=\frac{\pi a^2\sqrt{1-e^2}}{T}
\end{equation}
for the elliptical orbit, where we have used $S_{\rm ellipse}=\pi ab$ with $b=a\sqrt{1-e^2}$ the semi-minor axis.

Thus we get
\begin{equation}
\frac{\dot{\theta}'}{\dot{\theta}}=\frac{r_{\rm E}^2}{a^2}\frac{1}{\sqrt{1-e^2}}=\frac{(1-e^2)^{3/2}}{(1+e\cos\theta)^2}\simeq1-2e\cos\theta
\end{equation}
where the last approximation is valid to first order in $e$. Now, we have
\begin{equation}
\frac{{\rm d} \vartheta}{{\rm d}\theta}=\frac{{\rm d} \theta'}{{\rm d}\theta}-1=\frac{\dot{\theta}'}{\dot{\theta}}-1\simeq -2e\cos\theta
\end{equation}
and therefore
\begin{equation}
\vartheta= -2e\sin\theta.
\end{equation}
Since we are keeping only first-order terms in $e$, we can replace $\theta$ by $\theta'=2\pi(t-t_1)/T$ in the above equation, where $t_1$ is the time of perihelion passage. We finally get the following expression for $E_2$:
\begin{equation}
E_2(t)\simeq -\frac{de}{\pi}\sin\left(\frac{2\pi}{T}(t-t_1)\right).
\label{eq:e2}
\end{equation}
The dashed line on Fig.~\ref{fig:EOT}b shows $E_2(t)$. Combining (\ref{eq:e1}) and (\ref{eq:e2}), we find the expression (\ref{eq:EOT}) given in the text (solid line on Fig.~\ref{fig:EOT}b).


\begin{thebibliography}{99}
\bibitem{horizontality}
This could be improved easily, e.g. by using bubble levels to adjust the horizontality of the base.
\bibitem{footnote}
Note that with a fixed-base sundial, the determination of the equation of time can be much more accurate than with the method used in this paper. Indeed, in that case one needs to determine the time $t_{\rm max}$ at which the solar azimuth $\psi_\odot$ vanishes (i.e., when the Sun is South and crosses the observer's meridian), which can be done with a high precision as it varies linearly in time: the determination of such a zero-crossing is much more accurate than that of the position of a maximum.
\end{thebibliography}
\end{document}